\documentclass[12pt,english]{iopart}
\usepackage[english]{babel}
\usepackage{iopams} 
\usepackage{graphicx}

\def\Journal#1#2#3#4#5{{#1} #5 {\it #2} \textbf{#3} #4}
\def\Book#1#2#3#4#5{{#1} #5 {\it #2} (#3: #4)}

\newcommand{\EE}{\mathbb{E}}

\newcommand{\Pm}{\mathbb{P}}
\newcommand{\uu}{\mathbf 1}
\newcommand{\FF}{\mathcal{F}\/}
\newcommand{\GG}{\mathcal{G}\/}
\newcommand{\HH}{\mathcal{H}\/}
\newcommand{\CC}{\mathcal{C}\/}
\newcommand{\DD}{\mathcal{D}\/}
\newcommand{\ab}{a,b}

\newcommand{\dd}{{\rm d}}
\newcommand{\call}{P^{+}}
\newcommand{\putt}{P^{-}}
\newcommand{\caput}{P^{\pm}}
\newcommand{\bt}{t}

\begin{document}

\title[Perpetual American vanilla option pricing under single regime change risk]{Perpetual American vanilla option pricing under single regime change risk. An exhaustive study.}
\author{Miquel Montero}
\address{Departament de F\'{\i}sica Fonamental, Universitat de Barcelona,\\  Diagonal 647, E-08028 Barcelona, Spain.} 
\ead{miquel.montero@ub.edu}
\date{\today}

\begin{abstract}
Perpetual American options are financial instruments that can be readily exercised and do not mature. In this paper we study in detail the problem of pricing this kind of derivatives, for the most popular flavour, within a framework in which some of the properties |volatility and dividend policy| of the underlying stock can change at a random instant of time, but in such a way that we can forecast their final values. Under this assumption we can model actual market conditions because most relevant facts usually entail sharp predictable consequences. The effect of this potential risk on perpetual American vanilla options is remarkable: the very equation that will determine the fair price depends on the solution to be found. Sound results are found under the optics both of finance and physics. In particular, a parallelism among the overall outcome of this problem and a phase transition is established.
\end{abstract}

%\maketitle

\section{Introduction}
\label{S1}
Pricing financial derivatives is a main subject in mathematical finance with multiple implications in physics. In 1900, five years before Einstein's classic paper, Bachelier~\cite{C64} proposed the {\it arithmetic\/} Brownian motion for the dynamical evolution of stock prices with the aim of obtaining a formula for option valuation. Samuelson~\cite{PAS65} noticed the failure of such market model that allowed negative values in stock prices, and introduced the {\it geometric\/} Brownian motion which corrects this unwanted feature. Within his log-normal model, Samuelson obtained the fair price for perpetual options, although he was unable to find a general solution for expiring contracts. The answer to this question must wait until the publication of the works of Black and Scholes~\cite{BS73}, and Merton~\cite{M73}. The celebrated Black-Scholes-Merton formula has been broadly used by practitioners since then, mainly due to its unambiguous interpretation and mathematical simplicity. 

This mathematical simplicity in the Black-Scholes-Merton scheme has nevertheless a drawback:  
the model poorly adapts to those evolving conditions that affect actual derivatives and real markets present. In particular, empirical analyses conclude that volatility, roughly the diffusion parameter, must be considered as a changing (random) magnitude rather than a mere constant, as in the Black-Scholes-Merton formula. Many models have been developed in this direction, but among them a few deserve special emphasis because their historical imprint: the works of  Hull and White~\cite{HW87}, Wiggins~\cite{W87}, Scott~\cite{S87}, Stein and Stein~\cite{SS91}, or Heston~\cite{H93} belong to this selected group. They are collectively termed as {\it stochastic volatility\/} models because volatility becomes a continuous process that follows its own stochastic differential equation. Quite in the opposite direction we find another possible approach to the issue: the Markov-modulated geometrical Brownian motion models. Within these models the market coefficients change in a deterministic way but at random times, and according to that they are generically known as regime-switching models. Behind these lies the theory that the value of the parameters depends on the state of the economy, that suffers from seasonal changes. Seeds of this idea appeared in Barone-Adesi and Whaley~\cite{BAW87}, but the first time a model of this kind was properly settled is in the work of Naik~\cite{N93}. Since then such models have been used to discuss European~\cite{DMKR94,H98,MM04,ECS05}, Russian~\cite{G01} or American~\cite{BE02,GZ04,MM07,BL08,JP08} option prices, but also in asset allocation and portfolio optimisation problems~\cite{EVH97,ZY03,YZ04,SH04,ES08}.

Once we have mentioned different option flavours, we must also recall that the Black-Scholes-Merton formula is only is applicable to European options |derivatives that can be exercised  at maturity alone|, whereas most of the exchange-traded options are American |they can be exercised anytime during the life of the contract. Kim~\cite{IJK90} provided an integral representation of the American {\it vanilla\/} option price, but the explicit solution remains still unknown. 

The real difficulty behind pricing American options lies in the fact that one has to solve free boundary problems for partial differential equations, equations which sometimes have clear physical interpretation: McKean~\cite{McK65}, in what is the earliest approach to the issue, unambiguously speaks about the {\it heat equation\/} in an economic publication. Here the boundary represents the optimal exercise price, the stock price that triggers the early exercise of the option, and in general one can determine it just once the pricing expression is known, what leads to a circular problem. As a consequence, only restricted circumstances allow for closed formulas in American-like problems, whereas in the most general scenario analytical or numerical approximate methods must be used instead. One of theses favourable instances corresponds to a major simplification in the American problem: the assumption that the option never expires. Non-expiring or {\it perpetual\/} options serve as a good stating point in the resolution of the complete problem~\cite{BAW87} because the absence of maturity usually removes any explicit temporal dependence that those differential equations present.

In this article we are going to tackle the problem of pricing perpetual American vanilla calls and puts within a market model that can be considered as a degenerated instance of a regime-switching model with only two states and where one of the transition probabilities is set to zero: We will let volatility and dividend rate perform a {\it single\/} regime change at some unknown instant in the future. The value before and after the regime change of such magnitudes are assumed to be known in advance, what indeed determines if the change in the dynamics has taken place or not. 

Part of the present analysis is formally included as a limiting case in the paper by Guo and Zhang cited above~\cite{GZ04}, where the authors studied perpetual American vanilla put options with regime switching in absence of dividends, but there are divergences as well. On the one hand, we will also pay attention to the call case. Since dividend-paying assets are considered, a perpetual call may differ in price from its underlying, even though the option must not be early exercised. On the other hand, their study on puts is complete and very general but few explicit details about the closed-form solution found are given. This is specially relevant in the   analysis of the properties of the equations that determine the optimal exercise price: In essence one obtains different pricing formulas, each of them coming from an extreme value problem, and must elucidate the right choice at every moment. In spite of the multiplicity of parameters that our model involves, we will show how the value of a single magnitude answers the dilemma. Therefore, there is a formal resemblance between this behaviour and a phase transition in thermodynamics.

The paper is structured as follows: In \Sref{S2} we present the market model, the securities traded in it and their general properties. In \Sref{S3} we introduce the concept of hedging portfolio and show how it can be used for pricing derivatives. 
\Sref{S4} is specifically devoted to perpetual American vanilla options, we settle the properties of these ideal derivatives, and stress their interest from the point of view of finance and physics as well. In \Sref{S5} we quote the analytic expressions found, and emphasize the interpretation of the optimal exercise price as the outcome of an extremal problem. In \Sref{S6} we discuss the most appealing properties and particularities of the formulas that conform the solution to the problem, by illustrating our inferences through graphical examples. Conclusions are drawn in \Sref{Last}, and the paper ends with a lengthy Appendix where we present step-by-step calculations for call options. 

\section{The market}
\label{S2}
Let us begin with the general description of the stochastic properties of the securities that are traded in the market. 
The first security to be considered is a zero-coupon bond, a riskless monetary asset with a market price $B_t$ with deterministic evolution:
\begin{equation}
B_t=B_0 e^{r t}, \label{dB}
\end{equation}
where $r$, the risk-free interest rate, is assumed to be constant and positive. 

The second security present in the market is the stock $S_{t}$, whose time evolution for $t>0$ fulfils the following It\^o stochastic differential equation (SDE):   
\begin{equation}
\dd S_t=\mu_{t-} S_{t-} \dd t+ \sigma_{t-} S_{t-} \dd W_{t},\label{dS}
\end{equation}
where $W_{t}$ is a Wiener process, a one dimensional Brownian motion with zero mean and variance equal to $t$. The drift, $\mu_{t}$, and the volatility, $\sigma_{t}>0$, are stochastic processes as well: we will assume that their initial values are $\mu_a$ and $\sigma_a$, and after that moment they may simultaneously change to some other (different) fixed values, $\mu_b$ and $\sigma_b$, known in advance. Such a turnover can take place only once in a lifetime:~\footnote{Throughout the text $\uu_{\{\cdot\}}$ will denote the indicator function, which assigns the value 1 to a true statement, and the value 0 to a false statement.}
\begin{equation}
\mu_{t}=\mu_a \uu_{t< \tau} +\mu_b \uu_{t\geqslant\tau}=\mu_a +\left(\mu_b-\mu_a\right) \uu_{t\geqslant\tau},
\label{mu}
\end{equation}
\begin{equation}
\sigma_{t}=\sigma_a \uu_{t< \tau} +\sigma_b \uu_{t\geqslant\tau}=\sigma_a+\left(\sigma_b-\sigma_a\right)\uu_{t\geqslant\tau}.
\label{sigma}
\end{equation} 
We state that these magnitudes are stochastic because the instant $\tau>0$ in which the regime change occurs is a random variable. We will consider that $\tau$ %is exponentially distributed with 
follows an exponential law with $\EE^\Pm[\tau]=\lambda^{-1}>0$. In mathematical terminology both processes are {\it c\`adl\`ag\/} (right-continuous with left limits) semimartingales~\cite{K97,CT04}. In spite of the fact that $\mu_{t}$ and $\sigma_{t}$ are not continuous processes, the price itself is continuous because we are not considering any jump contribution in~(\ref{dS}). Therefore we may safely replace $S_{t-}$ by $S_{t}$ hereafter. %The question is not merely formal because here lies the core of the It\^o procedure of stochastic integration: the integrand must be {\it predictable\/}, in the sense that it cannot contain any random change to come. With this requirement in mind, we have decided to simplify the notation by dropping the explicit time arguments in SDE-related expressions.
Let us now analyse the meaning of~(\ref{mu}) and~(\ref{sigma}). The case of $\sigma_{t}$ is somewhat simpler since when $\sigma_a \neq \sigma_b$ we may speak of a (elementary) stochastic volatility market model with just two possible volatility levels. The meaning of the change in $\mu_{t}$ admits several interpretations~\cite{GZ04,MM07,SH04}, but we will concentrate our attention in the existence of two different (continuous-time) dividend pay-off regimes, $\mu_t=\tilde{\mu}_t-\delta_t$,
\begin{equation}
\delta_{t}=\delta_a \uu_{t< \tau} +\delta_b \uu_{t\geqslant\tau}=\delta_a+\left(\delta_b-\delta_a\right)\uu_{t\geqslant\tau},
\label{delta}
\end{equation}
with $\delta_{\ab}\geqslant 0$.  

A third kind of securities are negotiated in the market: the derivatives of $S$. Derivatives~\cite{S04} are financial instruments whose value depends on (derives from) present and past quotes of one or more underlying assets, which commonly are simply referred as {\it the underlyings\/}.
In particular, we are interested in %American 
options: contracts between two parties that give the buyer the right, but not the obligation, to buy (call) or sell (put) shares of the underlying stock at some prearranged price, the {\it strike price\/} $K>0$, at or before some specific instant in the future, the maturity or expiration time $T$. More precisely, we will assume that the option depends on the underlying by means of the present asset value alone, $S_t$. However, the option price does also depend on the stochastic process $I_t$, $I_t\equiv {\mathbf 1}_{t\geqslant \tau}$, through the market parameters $\tilde{\mu}_t$, $\sigma_t$ and $\delta_t$. The statement that an indicator function with a random variable in its argument constitutes a stochastic process may seem a delicate issue but it is mathematically correct: In fact, $I_t$ is a submartingale under our measure and, by virtue of the Doob-Meyer decomposition theorem~\cite{K97}, it can be expressed as a sum of two terms, $I_t=A_t +M_t$, the increasing adapted process $A_t$, 
\begin{equation*}
A_{t}=\lambda \min(\tau,t), 
%\label{Adef}
\end{equation*}
and the {\it c\`adl\`ag\/} (local) martingale $M_t$,
\begin{equation*}
M_{t}=1-\lambda \EE^\Pm[\tau|\FF_{t}]. 
%\label{Mdef}
\end{equation*}
In the previous expression $\FF_{t}$ represents all the available information up to time $t$, and corresponds to what is known in mathematical terminology as a filtration: an increasing one-parameter family of sub $\sigma$-algebras of $\FF$, $\FF_{t}\subseteq \FF_s \subseteq \FF$, $t \leqslant s$, where $\FF$ is a $\sigma$-algebra of subsets of the sample space $\Omega$. The sample space, $\FF$ and the measure $\Pm$ |which appeared beside the expectation symbol|  define the probability space ($\Omega,\FF,\Pm$). Finally, note that maturity introduces explicit temporal dependency in the value of the option as well, so we will have that $P_t\equiv P(t,S_t,I_t)$.

In the present case in which the price of the option is a function of two stochastic processes, one of them with discontinuous sample paths, the appropriate It\^o formula~\cite{CT04} reads:
\begin{eqnarray}
P(t,S_t,I_t)&=&P_a(0,S_0)+\int_0^t \partial_s P(s,S_s,I_{s-}) \dd s+\int_0^t \partial_{S} P(s,S_s,I_{s-}) \dd S_s \nonumber \\
&+&\frac{1}{2} \int_0^t \sigma_{s-}^2 S_{s}^2 \partial_{SS}^2 P(s,S_s,I_{s-}) \dd s+ \Delta P_\tau\ I_t,\label{ItoP}
\end{eqnarray}
where
\begin{equation*}
\Delta P_\tau \equiv P_b(\tau,S_\tau)-P_a(\tau,S_\tau),
\end{equation*}
and $P_b(t,S_t)=P(t,S_t,1)$, $P_a(t,S_t)=P(t,S_t,0)$. We can infer the discontinuous nature of the option price from the presence of the last term in~(\ref{ItoP}). The differential form of this expression is:     
\begin{equation}
\dd P_{t}=\partial_t P_{t-} \dd t+\partial_{S} P_{t-} \dd S_t+\frac{1}{2} \sigma_{t-}^2 S_{t}^2 \partial_{SS}^2 P_{t-} \dd t+ \Delta P_{t-} \dd I_t, \label{ItodP}
\end{equation}
where the property $P_{t-}\equiv P(t-,S_{t-},I_{t-})=P(t,S_t,I_{t-})$ has been systematically used. Depending on the notation used, one can find that the $\Delta P_{t-} \dd I_t$ term from the above expression is simply replaced with $\Delta P_{t-}$, but then one must define the random measure associated to the jumping process. We have decided not to proceed in that way because the appearance of $\dd I_t$ in the forthcoming expressions stresses the option price dependence on a stochastic process different from $\dd S_t$. % not related with $\dd W_t$, which does not contribute in an explicit way to the variation of the stock price $\dd S_t$ |{\it i.e.\/}, it does not appear in~(\ref{dS}). 
Therefore, it is a source of risk that cannot be explained in terms of the random evolution of the underlying asset. When part of the risk is not directly traded in the market, the market may be incomplete: we will not be able to reproduce the behaviour of the option by means of a replicating or {\it hedging\/} portfolio. 
The immediate consequence of such eventuality is that the risk premium coming from $\tau$ is arbitrary to a certain extent, because investors can evaluate it on the basis of their own perceptions. We will see all of this in more detail in the next Section.

\section{Option pricing}
\label{S3}
In this Section we will obtain pricing expressions for the derivatives of $S$ under the market conditions specified above, by resorting to the hedging-portfolio technique. This approach is based on the idea that the fair price of an option must be equal to the value of some portfolio $\Pi$ made of different securities that mimics the behaviour of the derivative and hedges all the risk. If such a portfolio can be defined, the market model does not admit arbitrage~\cite{S04}. 

The first security to be included in the portfolio is the underlying asset, which will reproduce changes in the option price due to the evolution of the stock price $S$. The second security in our portfolio is $B$: a long position in this security will provide a secure resort where to keep the benefits of an effective trading strategy, whereas a short position in these bonds will allow us to borrow money when we need it. 
These two securities cannot counterbalance all the stochastic behaviour of the price of the option as we have pointed out before: not all the influence of $\delta_{t}$ and $\sigma_{t}$ on the option price may be explained through $S_t$. Therefore, we need another security that can account for this contribution to the global risk. Since in the most usual situation markets do not trade such assets, we will consider the inclusion of a {\it secondary\/} option $Q$ in the portfolio: a derivative of the same nature of $P$, but with a different set of contract specifications. In particular we will assume that $Q$ has a striking price $K'$, different from $K$. This is a standard technique~\cite{BBR96,RT97} intended to complete the market.

Summing up, our aim is to reproduce the evolution of $P_t$ by means of a portfolio $\Pi_t$ built out of a mixture of $\nu_t$ shares $S_t$, $\phi_t$ units of the riskless security $B_t$, and $\psi_t$ secondary options with value $Q_t$:
\begin{equation}
\Pi_t=\nu_t S_t + \phi_t B_t + \psi_t Q_t,
\label{Pdef}
\end{equation}
in such a way that $\Pi_t= P_t$ holds almost surely. The predictable {\it c\`agl\`ad\/} (left-continuous with right limits) processes $\nu_t$, $\phi_t$ and $\psi_t$ constitute what is known in mathematical finance terminology as a trading strategy, and the capital gain $\DD_t$ associated to it is given by:
\begin{equation*}
\DD_t=\int_0^t \nu_t \dd S_t + \int_0^t \phi_t \dd B_t + \int_0^t \psi_t \dd Q_t + \int_0^t \nu_t \delta_{t-} S_t \dd t,
\end{equation*}
where the last term represents the amount of money received as dividends. The cost process $\CC_t$ associated to the strategy is just the difference between the value of the portfolio $\Pi_t$ and the revenues $\DD_t$, $\CC_t=\Pi_t-\DD_t$, and the trading strategy is said to be {\it self-financing\/} if $\CC_t=0$, because in that case there is no net cash flow entering or leaving the replicating portfolio~\cite{CT04}. When $\CC_t=0$ the variation in the value of the portfolio $\dd \Pi_t$ coincides with the change in the capital gain $\dd \DD_t$:
\begin{equation}
\dd \Pi_t=\nu_t \dd S_t + \phi_t \dd B_t + \psi_t \dd Q_t + \nu_t \delta_{t-} S_t \dd t.
\label{dP}
\end{equation}

Now we have to demand to that SDE~(\ref{dP}) matches~(\ref{ItodP}):
\begin{eqnarray}
\partial_t P_{t-} \dd t+\partial_{S} P_{t-} \dd S_t+\frac{1}{2} \sigma^2_{t-} S_t^2 \partial_{SS}^2 P_{t-} \dd t+ \Delta P_{t-} \dd I_t= \nonumber \\
\nu_t \dd S_t + \phi_t \dd B_t + \psi_t \dd Q_t+ \nu_t \delta_{t-} S_t \dd t. 
\label{PDE_1}
\end{eqnarray}
In an analogous way to $\dd P_t$, $\dd Q_t$ follows a SDE,
\begin{equation}
\dd Q_t=\partial_t Q_{t-} \dd t+\partial_{S} Q_{t-} \dd S_t+\frac{1}{2} \sigma^2_{t-} S^2_t \partial_{SS}^2 Q_{t-} \dd t+ \Delta Q_{t-} \dd I_t, \label{dQ}
\end{equation}
where 
\begin{equation*}
\Delta Q_t=Q(t,S_t,1)-Q(t,S_t,0)=Q_b(t,S_t)-Q_a(t,S_t).
\end{equation*}
We may now combine~(\ref{dB}) and~(\ref{dQ}) with equation~(\ref{PDE_1}) and get:
\begin{eqnarray}
\Bigg[\partial_t P_{t-} +\frac{1}{2} \sigma^2_{t-} S^2_t \partial_{SS}^2 P_{t-}-\nu_t \delta_{t-} S_t -r \phi_t B_t \nonumber \\
-\psi_t\left(\partial_t Q_{t-} + \frac{1}{2} \sigma^2_{t-} S^2_t \partial_{SS}^2 Q_{t-} \right) \Bigg]\dd t= \nonumber \\
\left(\nu_t -\partial_{S} P_{t-}+\psi_t \partial_{S} Q_{t-}\right)\dd S_t - \left(\Delta P_{t-} -\psi_t \Delta Q_{t-}\right)\dd I_t. 
\label{PDE_2}
\end{eqnarray}
In order to transform~(\ref{PDE_2}) from a SDE into a deterministic partial differential equation we must guarantee that terms containing the stochastic magnitudes $\dd S_t$ and $\dd I_t$ cancel out, by selecting the self-financing trading strategy. Therefore, we must demand that
\begin{equation*}
\nu_t=\partial_{S} P_{t-} - \psi_t \partial_{S} Q_{t-},
\end{equation*}
a condition named {\it delta hedging\/}, and also that
\begin{equation*}
\psi_t=\frac{\Delta P_{t-}}{\Delta Q_{t-}},
\end{equation*}
which is usually referred as {\it vega hedging\/}. The previous hedging conditions reduce~(\ref{PDE_2}) to
\begin{eqnarray}
\partial_t P_{t-} +\frac{1}{2} \sigma^2_{t-} S^2_t \partial_{SS}^2 P_{t-} -\delta_{t-} S_t \partial_{S} P_{t-} = \nonumber \\
  r \phi_t B _t + \frac{\Delta P_{t-}}{\Delta Q_{t-}} \left( \partial_t Q_{t-} +\frac{1}{2} \sigma^2_{t-} S^2_t \partial_{SS}^2 Q_{t-} -\delta_{t-} S_t \partial_{S} Q_{t-} \right), \label{CBnD}
\end{eqnarray}
an expression that still involves $\phi_t B_t$. We can make it disappear just by using the definition of the portfolio in~(\ref{Pdef}), the predictable character of $\phi_t$, the delta hedging and the vega hedging, altogether,
\begin{eqnarray*}
\phi_t B_t&=&P_{t-}-\nu_t S_t - \psi_t Q_{t-}\\
&=&P_{t-} -\left(\partial_{S} P_{t-} - \frac{\Delta P_{t-}}{\Delta Q_{t-}}\partial_{S} Q_{t-} \right)S_t - \frac{\Delta P_{t-}}{\Delta Q_{t-}} Q_{t-}.
\end{eqnarray*}
The replacement of $\phi_t B_t$ in~(\ref{CBnD}) leads to
\begin{eqnarray*}
\partial_t P_{t-} +\frac{1}{2} \sigma^2_{t-} S^2_t \partial_{SS}^2 P_{t-} -r P_{t-} +(r-\delta_{t-}) S_t\partial_S P_{t-}
=\\
\frac{\Delta P_{t-}}{\Delta Q_{t-}} \left[ \partial_t Q_{t-}+\frac{1}{2} \sigma^2_{t-} S^2_t \partial_{SS}^2 Q_{t-}-r Q_{t-} +(r-\delta_{t-})S_t\partial_S Q_{t-}\right].
\end{eqnarray*}
This formula implies the existence of a generic predictable process $\Xi_t=\Xi(t,S_t,I_{t-})$~\cite{H98,MM04}, which uncouples the problem of finding $P_t$ and $Q_t$:
\begin{equation}
\partial_t P_{t-} +\frac{1}{2} \sigma^2_{t-} S^2_t \partial_{SS}^2 P_{t-} -r P_{t-} +(r-\delta_{t-}) S_t\partial_S P_{t-}+\Xi_{t} \Delta P_{t-}=0,
\label{Xi}
\end{equation}
and proves that the option $Q_t$ removes the risk associated with $I_t$. %and therefore completes the market.
Formula~(\ref{Xi}) is valid for the secondary option as well, merely by replacing $P_t$ with $Q_t$. Note that, whereas~(\ref{sigma}) and~(\ref{delta}) affect this expression, it contains no reference either on $\tilde{\mu}_t$ or $\lambda$. Before discussing the financial interpretation of $\Xi_t$, let us briefly analyse formula~(\ref{Xi}). As before we will use the notation $\Xi_a(t,S_t)\equiv \Xi(t,S_t,0)$ and $\Xi_b(t,S_t)\equiv \Xi(t,S_t,1)$. When $t<\tau$ the differential equation~(\ref{Xi}) reduces to  
\begin{equation}
\partial_t P_a +\frac{1}{2} \sigma^2_a S^2_t \partial_{SS}^2 P_a -r P_a +(r-\delta_a) S_t\partial_S P_a+\Xi_a \left(P_b-P_a\right)=0,
\label{Xi_a}
\end{equation}
whereas for $t>\tau$ we will have
\begin{equation*}
\partial_t P_b +\frac{1}{2} \sigma^2_b S^2_t \partial_{SS}^2 P_b -r P_b +(r-\delta_b) S_t\partial_S P_b+\Xi_b \left(P_b-P_a\right)=0.
%\label{Xi_b}
\end{equation*}
Note however that in this case $P_b$ must fulfil the classical Black-Scholes-Merton equation with continuous-time dividends, {\it i.e.\/}:
\begin{equation}
\partial_t P_b +\frac{1}{2} \sigma^2_b S^2_t \partial_{SS}^2 P_b -r P_b +(r-\delta_b) S_t\partial_S P_b=0,
\label{Xi_b}
\end{equation}
therefore we have a first constraint for $\Xi_t$, $\Xi_b(t,S_t)=0$. Finally, since $P_t$ is a right-continuous stochastic process, we have that $P(\tau,S_\tau,I_\tau)=\lim_{t\downarrow \tau} P_b(t,S_t)$.

We will concentrate now on $\Xi_t$. To this end, let us introduce another portfolio $\Psi_t$, 
\begin{equation*}
\Psi_t \equiv \tilde \nu_t S_t +\tilde  \phi_t B_t + P_t,
\end{equation*}
based upon the following self-financing trading strategy
\begin{eqnarray*}
\tilde \nu_t&=&-\partial_{S} P_{t-},\\
\tilde \phi_t &=&\frac{S_t \partial_{S} P_{t-}-P_{t-}}{B_t}.
\end{eqnarray*}
On the one hand we have that $\Psi_{t-}=0$, and on the other hand
\begin{eqnarray*}
\dd \Psi_t &=& \tilde \nu_t \dd  S_t +\tilde  \phi_t\dd  B_t + \dd  P_t+\tilde \nu_t \delta_{t-} S_t \dd t\\
&=& -\partial_{S} P_{t-}\dd  S_t+ \left[(r-\delta_{t-}) S_t \partial_{S} P_{t-}-r P_{t-} \right]\dd t+\dd P_t \\
&=&\left[\partial_t P_{t-} +\frac{1}{2} \sigma^2_{t-} S^2_t \partial_{SS}^2 P_{t-} -r P_{t-} +(r-\delta_{t-}) S_t\partial_S P_{t-}\right]\dd t+\Delta P_{t-}\dd I_t\\
&=&\Delta P_{t-}\left( \dd I_t - \Xi_{t}\dd t\right),
\end{eqnarray*}
where we have used~(\ref{ItodP}) and~(\ref{Xi}). Let us analyse now the right-hand side of the last equality in the previous equation for $t\leqslant\tau$, and assume that for some range of values of its arguments $\Xi_a(t,S_t)<0$, and therefore $\Xi_{t}<0$. Moreover, let us restrict our considerations to a subset in which we have simultaneously $\Xi_{t}< 0$ and $\Delta P_{t-}>0$ |note that both quantities are predictable. The only thing we ought to do in order to obtain what is known as a ``free lunch'', {\it i.e.\/} sure earnings without capital exposure, is to wait until convenient market conditions are reached and costless construct $\Psi_{t-}$ immediately, because then $\dd \Psi_t> 0$. The possibility of performing this kind of investment is named {\it arbitrage opportunity\/}, and it constitutes an undesirable feature of any market model. Even if $\Xi_a=0$ we would get a ``free ticket'' for a lottery, $\dd \Psi_t>0$ with probability $\lambda \dd t$ |the regime change either takes place| and $\dd \Psi_t=0$ with probability $1-\lambda \dd t$ |or not. Note that if $\Delta P_{t-}<0$ the trick is to compose the reverse portfolio. Finally, the conclusion is valid for $\Delta P_{t-}=0$ as well, because $P(t,S_t,I_t)$ and $\Xi(t,S_t,I_{t-})$ are continuous in $t$ and $S_t$. 
$\Xi_a(t,S_t)>0$ is the only strong constraint on this magnitude, which depends in essence on how every investor measures volatility and dividend risks, that are not traded in the market. 

Since $\Xi_a$ cannot be unambiguously settled, the market is not complete strictly speaking. Therefore one must resort to some sound financial criteria to choose $\Xi_a$. A conventional choice is the one that removes the so-called statistical arbitrage that appears when the growth of the expected value of a null portfolio is different from zero. In our case this is attained under the measure $\Pm$ if $\Xi_a=\lambda$, since then for $t\leqslant \tau$ we have
\begin{eqnarray*}
\EE^\Pm\left[\left.\dd \Psi_t\right|\FF_{t-}\right]= \left(\lambda - \Xi_a \right)\Delta P_{t-}\dd t=0.
\end{eqnarray*}
%and therefore the partial differential equation to be solved reduces to
%\begin{equation}
%\partial_t P +\frac{1}{2} \sigma^2 S^2 \partial_{SS}^2 P -rP +(r-\delta) S\partial_S P+\lambda  P=-\lambda P_b.
%\label{PDE_t}
%\end{equation}

\section{Perpetual American vanilla options}
\label{S4}
The main assumption that determines the range of validity of equation~(\ref{Xi}) is the  irrelevance of past values of the underlying in the valuation of the derivative. This constraint is satisfied for any option whose value at the exercise time, $t^*$, depends only on the present price of the underlying, $S_{t^*}$, and $t^*$ itself:
\begin{equation*}
P(t^*,S_{t^*},I_{t^*})=X\left(S_{t^*}\right) \uu_{t^*\leqslant T}.
\end{equation*}
When the option can be exercised at the end of the contract lifetime only, $t^*=T$, the exercise time is deterministic, and the option is said to be European. If the option can be exercised at any time before expiration it is called American, and $t^*$ becomes a stochastic magnitude as well. Note that the contract is always worthless after maturity, $t^*>T$, and therefore the option buyer must decide under which conditions the option can be optimally exercised before this deadline. Since the decision may be based on %the present value of stock price 
$S_t$ and the natural time variable, the time to maturity, $T-t$, 
one deduces that the optimal exercise boundary, the stock price for which it is better to exercise the option than to keep it alive, will be of the form $H_t=H(t,I_t)$, with the usual auxiliary expressions $H_a(t)=H(t,0)$ and $H_b(t)=H(t,1)$ assumed. 
The very definition of $H_t$ entails that for any live American option it holds that
\begin{equation}
P(\bt,S_{\bt},I_{\bt}) \geqslant X\left(S_{\bt}\right), \label{better_alive}
\end{equation}
and that $H_{\bt}$ must be settled in such a way that relationship
\begin{equation*}
P(\bt,H_{\bt},I_{\bt})=X\left(H_{\bt}\right).
%\label{continuity}
\end{equation*}
is satisfied. As a consequence, the computation of the option price and the optimal exercise boundary must be tackled contemporarily. Here lies the root of the difficulty in pricing American derivatives: first one must solve the partial differential equation~(\ref{Xi}) for any possible boundary condition, $P(\bt,S_{\bt},I_{\bt}\ ;G_{\bt})$, 
\begin{equation*}
P(\bt,G_{\bt},I_{\bt}\ ;G_{\bt})=X\left(G_{\bt}\right),
\end{equation*}
and then infer the value of $H_{\bt}$ from some other financial argument. Note that option holders try to delay the exercise of the option as much as they can, since the worth of a live option is never lower than the pay-off of the contingent claim,~(\ref{better_alive}). Then $H_{\bt}$ is the function that maximizes the price of the option with respect to the boundary condition:  
\begin{equation}
P(\bt,S_{\bt},I_{\bt}\ ;H_{\bt})\geqslant P(\bt,S_{\bt},I_{\bt}\ ;G_{\bt}).
\label{max}
\end{equation}
When the pay-off function is regular enough, the above demand is equivalent to the classical smooth pasting condition that states~\cite{PAS65,M73}:
\begin{equation}
%\left. \frac{\partial P(\bt,S\ ;H_{\bt})}{\partial S}\right|_{S=H_{\bt}}=\left. \frac{\dd X(S)}{\dd S}\right|_{S=H_{\bt}}.
\left.\partial_S P(\bt,S,I_{\bt}\ ;H_{\bt})\right|_{S=H_{\bt}}=\left. \frac{\dd X(S)}{\dd S}\right|_{S=H_{\bt}}.
\label{smooth}
\end{equation}
In conclusion, the option price must be continuous with continuous derivative in the asset price when it crosses the boundary.

Few closed solutions for this general problem are known even when the stock price follows a standard geometric Brownian motion. In particular the question stands open for the most ubiquitous pay-off function, 
\begin{equation}
X^{\pm}(S)=\max\left(\pm(S-K),0\right).
\label{Vpo}
\end{equation}
This kind of derivatives are known as {\it vanilla\/} options, and the plus (respectively minus) superscript stands for call (respectively put) options. The difficulty deceases dramatically if one can deduce the way in which the optimal exercise boundary depends on time. This is the case when the option is perpetual, when it does not mature. It can be objected that perpetual American options have limited practical interest since actual derivatives commonly expire. However, since they can be understood as an approximation of a far-from-maturity contract, perpetual options may help in the valuation process when theoretical prices cannot be computed~\cite{BAW87}. Moreover, as we will stress later, the emerging problem is still akin to physics because it involves mechanical or thermodynamic concepts like equilibrium and stability.

We will translate now the absence-of-expiration assumption into our expressions by letting $T\rightarrow \infty$. %or, what is the same, $\bt \rightarrow \infty$. 
This will remove any {\it explicit\/} temporal dependence from $P_t$, $P(t,S_t,I_t) \rightarrow P(S_t,I_t)$ and from $H_{\bt}$,  $H(t,I_t) \rightarrow H(I_t)$. So, the remaining {\it implicit\/} time dependence in $H_t$ is fruit of a possible future change in the dynamics exclusively, and therefore 
\begin{equation*}
H_{t}=H_a \uu_{t< \tau} +H_b \uu_{t\geqslant\tau},
\end{equation*}
with $H_a$ and $H_b$ non-negative constants to be determined.

\section{Main analytic results}
\label{S5}
In this Section we will present the solution to equations~(\ref{Xi_a}) and~(\ref{Xi_b}), with $\Xi_a=\lambda$, when the pay-off function is~(\ref{Vpo}). The detailed derivation of the expressions to appear can be found in~\ref{AA} for the case of call options, and one may use the procedure developed there as a guideline for finding put formulas.

After the regime change we will have~\cite{BAW87}:~\footnote{We will drop the subscript in $S_t$ hereafter for notational brevity.}
\begin{equation*}
\caput_b(S;H^{\pm}_b)=\pm(H^{\pm}_b - K) \left(\frac{S}{H^{\pm}_b}\right)^{\beta^{\pm}_b}, (S\lesseqgtr H^{\pm}_b),
\end{equation*}
with
\begin{equation*}
H^{\pm}_b = \frac{\beta^{\pm}_b}{\beta^{\pm}_b - 1} K. \label{Hb}
\end{equation*}
Here and hereafter we will denote by $\beta^{\pm}_{\ab}$ the following combination of the model parameters $r$, $\delta_{\ab}$, and $\sigma_{\ab}$:
\begin{equation}
\beta^{\pm}_{\ab} = \frac{1}{\sigma^2_{\ab}}\left(-\theta_{\ab} \pm \sqrt{\theta^2_{\ab}+2 r \sigma^2_{\ab}}\right),
\label{betapm}
\end{equation}
where $\theta_{\ab}$ is a shorthand for
\begin{eqnarray*}
\theta_{\ab}=r-\delta_{\ab}-\sigma_{\ab}^2/2.
\end{eqnarray*}
Note that $\beta^{\pm}_{\ab}\gtrless 0$. In fact, one can prove that $\beta^{+}_{\ab}\geqslant 1$ and $0>\beta^{-}_{\ab}\geqslant -2r/\sigma_{\ab}^2$, where equalities hold if and only if $\delta_{\ab}=0$.

The relative values of $\beta^{\pm}_a$ and $\beta^{\pm}_b$ play a key role in the evaluation of the price of the perpetual option before the regime change. When $|\beta^{\pm}_a|\geqslant |\beta^{\pm}_b|$, {\it i.e.\/} when $\beta^{\pm}_a\gtreqless \beta^{\pm}_b$, one finds that necessarily $H^{\pm}_a\lesseqgtr H^{\pm}_b$, and the value of $\caput_a(S;H^{\pm}_a)$ can be encoded in a single expression: 
\begin{eqnarray}
\caput_a (S;H^{\pm}_a)=\pm
(H^{\pm}_a - K)\left(\frac{S}{H^{\pm}_a}\right)^{\gamma^{\pm}_{a}} \nonumber \\  \pm(H^{\pm}_b - K)\frac{\lambda}{\lambda+\ell^{\pm}} \left[\left(\frac{S}{H^{\pm}_b}\right)^{\beta^{\pm}_b} -\left(\frac{H^{\pm}_a}{H^{\pm}_b}\right)^{\beta^{\pm}_b} \left(\frac{S}{H^{\pm}_a}\right)^{\gamma^{\pm}_{a}} \right], (S\lesseqgtr H^{\pm}_a),
\label{Vcaputa1}
\end{eqnarray}
where we have introduced $\gamma^{\pm}_{a}\gtrless 0$ which incorporates $\lambda$:
\begin{equation}
\gamma^{\pm}_{a}=\frac{1}{\sigma^2_a}\left(-\theta_a \pm \sqrt{\theta^2_a+2 (r+\lambda) \sigma^2_a}\right),
\label{gamma}
\end{equation}
and constant $\ell^{\pm}$ which depends on the remaining parameters of the market model:
\begin{eqnarray}
\ell^{\pm} &=& \left(\beta^{\pm}_a-\beta^{\pm}_b\right)\left(\frac{r}{\beta^{\pm}_a}+\frac{1}{2}\sigma^2_a\beta^{\pm}_b\right), 
\label{ell}
\end{eqnarray}
and cancels if and only if $\beta^{\pm}_a=\beta^{\pm}_b$.
The value of the optimal exercise price cannot be obtained in a closed form valid for any value of the parameters. As we have explained above one can recover $H^{\pm}_a$ by demanding the so-called smooth pasting condition~(\ref{smooth}) on the solution~(\ref{Vcaputa1}). This one is the approach we follow in the Appendix because it naturally provides ordinary differential equations with Cauchy boundary conditions on $S=H^{\pm}_a$. Here we will show how~(\ref{smooth}) is just a consequence of the general maximal principle~(\ref{max}) which in this case takes the form:
\begin{equation}
\frac{\partial \caput_a(S;H^{\pm}_a)}{\partial H^{\pm}_a}=0.
\label{maxH}
\end{equation}
If one differentiates~(\ref{Vcaputa1}) with respect to $H^{\pm}_a$ gets
\begin{equation*}
\frac{\partial \caput_a(S;H^{\pm}_a)}{\partial H^{\pm}_a}=\mp \gamma^{\pm}_a \frac{\beta^{\pm}_b-1}{\beta^{\pm}_b} \left(\frac{S}{H^{\pm}_a}\right)^{\gamma^{\pm}_{a}}\left(\frac{H^{\pm}_a}{H^{\pm}_b}\right)^{\beta^{\pm}_b-1} \GG^{\pm}\left(\frac{H^{\pm}_a}{H^{\pm}_b}\right),
\end{equation*}
where we have defined the auxiliary function $\GG^{\pm}(\eta)$, 
\begin{equation*}
\GG^{\pm}(\eta) = A^{\pm} \eta^{1-\beta^{\pm}_b} -\eta^{-\beta^{\pm}_b} \mp B^{\pm}, 
\label{Gpm}
\end{equation*}
and the constant positive-definite coefficients~\footnote{One can consider different definitions for the auxiliary function and the associated coefficients. We kept the same notation appearing in~\cite{MM07}.}
\begin{eqnarray}
A^{\pm}&=&  \frac{\gamma^{\pm}_a-1}{\gamma^{\pm}_a}\frac{\beta^{\pm}_b}{\beta^{\pm}_b-1}, \label{cA} \\
B^{\pm}&=& \frac{\mp 1}{\beta^{\pm}_b-1}\frac{\lambda}{\lambda+r}\frac{\gamma^{\mp}_a}{\beta^{\pm}_b-\gamma^{\mp}_a}. \label{cB}
\end{eqnarray}
Therefore we have to look for $\eta^{\pm}_0$, $\GG^{\pm}(\eta^{\pm}_0)=0$, because then the value of $H^{\pm}_a$ will follow,
\begin{equation*}
H^{\pm}_a =\eta^{\pm}_0 H^{\pm}_b.
\end{equation*}
It is shown in the Appendix that function $\GG^{+}(\eta)=0$ has one and only one zero in the interval $0<\eta\leqslant 1$, but the same kind of analysis performed there leads to the conclusion that $\GG^{-}(\eta)=0$ has a single solution in  $1\leqslant\eta <\infty$. Thus the problem is properly settled.  

When $|\beta^{\pm}_a|< |\beta^{\pm}_b|$ the optimal exercise price for vanilla calls (respectively puts) is higher (respectively lower) before the regime change than after it. This makes feasible that a change in the dynamics compels the holder to early exercise the option immediately because the sudden change in the optimal boundary value. This fact leads to the existence of two different pricing formulas depending on the present value of the underlying: 
\begin{eqnarray}
\caput_a(S;H^{\pm}_a)=\pm\Bigg[\frac{\delta_a}{\lambda+\delta_a}H^\pm_a - \frac{r}{\lambda+r}K\Bigg]\left(\frac{S}{H^\pm_a}\right)^{\gamma^\pm_a}\nonumber \\
\pm\frac{\lambda}{\lambda+r}\frac{K}{\gamma^{\pm}_a-\gamma^{\mp}_a} \Bigg\{\frac{\gamma^\mp_a}{\gamma^\pm_a-\beta^\pm_b}\Bigg[\frac{\beta^\pm_b}{\gamma^\pm_a-1} \left(\frac{S}{H^\pm_b}\right)^{\gamma^\pm_a}\nonumber \\
-\frac{\gamma^\pm_a}{\beta^\pm_b-1}\frac{\gamma^{\pm}_a-\gamma^{\mp}_a}{\beta^{\pm}_b-\gamma^{\mp}_a}\left(\frac{S}{H^{\pm}_b}\right)^{\beta^{\pm}_b} \Bigg]\nonumber \\
-\frac{\gamma^\pm_a\beta^\pm_b}{(1-\gamma^\mp_a)(\beta^\pm_b-\gamma^\mp_a)}\left(\frac{H^\pm_b}{H^\pm_a}\right)^{(\gamma^\pm_a-\gamma^\mp_a)}\left(\frac{S}{H^\pm_b}\right)^{\gamma^\pm_a}\Bigg\},  \quad (S\lesseqgtr H^{\pm}_b), 
\label{Vcaputa2}
\end{eqnarray}
and
\begin{eqnarray}
\caput_a(S;H^{\pm}_a)=\pm\left[\frac{\delta_a}{\lambda+\delta_a}H^\pm_a - \frac{r}{\lambda+r}K\right]\left(\frac{S}{H^\pm_a}\right)^{\gamma^\pm_a}\nonumber\\
\pm\frac{\lambda K}{\lambda+r}\frac{\gamma^\pm_a\beta^\pm_b}{(\gamma^{\pm}_a-\gamma^{\mp}_a)(1-\gamma^\mp_a)(\beta^\pm_b-\gamma^\mp_a)}\Bigg[\left(\frac{S}{H^\pm_b}\right)^{\gamma^\mp_a}-\left(\frac{H^\pm_a}{H^\pm_b}\right)^{\gamma^\mp_a}\left(\frac{S}{H^\pm_a}\right)^{\gamma^\pm_a}\Bigg]\nonumber\\
\pm\frac{\lambda}{\lambda+\delta_a}S \mp \frac{\lambda}{\lambda+r}K, \quad ( H^{\pm}_b\lessgtr S\lesseqgtr H^{\pm}_a). \label{Vcaputa3}
\end{eqnarray}
Both expressions match when $S=H^\pm_b$, in a very smooth way: one has to compute the fourth derivative of the option price with respect to the spot price before a discrepancy is found. The value of $H^\pm_a$ remains elusive: the optimal exercise price solves an equation that in the general situation is transcendental again. In this case the use of the maximization criterion~(\ref{maxH}) instead of the matching condition~(\ref{smooth}) is very illustrative as well. Since~(\ref{maxH}) is a global condition, it affects the whole solution. Then one may perform the computation either on~(\ref{Vcaputa2}) or on~(\ref{Vcaputa3}) with the same result, namely,
\begin{equation*}
\frac{\partial \caput_a(S;H^{\pm}_a)}{\partial H^{\pm}_a}=\pm \gamma^{\pm}_a \frac{\beta^{\pm}_b-1}{\beta^{\pm}_b}\frac{r}{\lambda+r} \left(\frac{S}{H^{\pm}_a}\right)^{\gamma^{\pm}_{a}}\left(\frac{H^{\pm}_a}{H^{\pm}_b}\right)^{\gamma^{\mp}_a-1} \HH^{\pm}\left(\frac{H^{\pm}_a}{H^{\pm}_b}\right),
\end{equation*}
with
\begin{eqnarray*}
\HH^\pm(\chi)=\chi^{-\gamma^\mp_a}\mp C^\pm \chi^{1-\gamma^\mp_a}\pm D^\pm,
\end{eqnarray*}
and
\begin{eqnarray}
C^{\pm}&=&\mp \frac{\gamma^\mp_a}{1-\gamma^\mp_a}\frac{\beta^{\pm}_b}{\beta^{\pm}_b-1}\frac{\delta_a}{r}, \label{cC}\\
D^{\pm}&=& \pm\frac{\lambda}{r}\frac{\beta^{\pm}_b}{(\beta^{\pm}_b-\gamma^\mp_a)(1-\gamma^\mp_a)}\label{cD}, 
\end{eqnarray}
positive magnitudes. As before, there is a single solution to the equation $\HH^\pm(\chi^{\pm}_0)=0$ that belongs to the interval $1<\chi^{\pm1}<\infty$. After obtaining the value of $\chi^{\pm}_0$ we can compute $H^{\pm}_a$ though
\begin{equation*}
H^{\pm}_a =\chi^{\pm}_0 H^{\pm}_b.
\end{equation*}

Note finally that in expression~(\ref{Vcaputa2}) it was assumed that $\beta^{\pm}_b\neq \gamma^{\pm}_a$. When a numerical concordance in the parameters leads to $\beta^{\pm}_b=\gamma^{\pm}_a$, we have merely to take the limit $\beta^{\pm}_b \rightarrow\gamma^{\pm}_a$ in the corresponding expression and the right solution is obtained: 
\begin{eqnarray}
\caput_a(S;H^{\pm}_a)=\pm\Bigg[\frac{\delta_a}{\lambda+\delta_a}H^\pm_a - \frac{r}{\lambda+r}K\Bigg]\left(\frac{S}{H^\pm_a}\right)^{\gamma^\pm_a}\nonumber \\
\pm\frac{\lambda}{\lambda+r}\frac{K}{\gamma^{\pm}_a-\gamma^{\mp}_a} \Bigg\{\frac{\gamma^\mp_a}{\gamma^\pm_a-1}\Bigg[\frac{1-2\gamma^\pm_a}{\gamma^\pm_a-1}-\frac{\gamma^{\pm}_a}{\gamma^{\pm}_a-\gamma^{\mp}_a} +\gamma^\pm_a \ln \left(\frac{S}{H^{\pm}_b}\right) \Bigg]\left(\frac{S}{H^\pm_b}\right)^{\gamma^\pm_a}\nonumber \\
-\frac{(\gamma^\pm_a)^2}{(1-\gamma^\mp_a)(\gamma^\pm_a-\gamma^\mp_a)}\left(\frac{H^\pm_b}{H^\pm_a}\right)^{(\gamma^\pm_a-\gamma^\mp_a)}\left(\frac{S}{H^\pm_b}\right)^{\gamma^\pm_a}\Bigg\},  \quad (S\lesseqgtr H^{\pm}_b). 
\label{Vcaputa4}
\end{eqnarray}
The question is that the case in which $\beta^{\pm}_b=\gamma^{\pm}_a$ does not constitute a critical point for the process, it only affects the functional form of the solution. There are some other situations with deeper financial relevance that deserve a closer study. We will introduce and illustrate them in the next Section. 

\section{Discussion}
\label{S6}

Let us analyse in the first place the consequences on call option prices of a stoppage in the payment of dividends, $\delta_a\neq0$ and $\delta_b=0$. After the change we have $\beta^{+}_b=1$, $H^{+}_b\rightarrow \infty$, and $\call_b\rightarrow S$. This ensures that the right choice here is the solution for which $H^{+}_a\leqslant H^{+}_b$. So we have to write equation~(\ref{Vcaputa1}) for the case in which $\beta^{+}_b=1$:
\begin{equation*}
\call_a(S;H^+_a)= \left[\frac{\delta_a}{\lambda +\delta_a}H^+_a-K\right] \left(\frac{S}{H^+_a}\right)^{\gamma^{+}_{a}}+\frac{\lambda}{\lambda +\delta_a}S, \quad (S\leqslant H^+_a),
\end{equation*}
where we have taken into account that here $\ell^{+}=\delta_a$~\cite{MM07}. Since $\beta^{+}_b\rightarrow 1$ implies $A^+\rightarrow \infty$ and $B^+\rightarrow \infty$, it is somewhat simpler to recompute~(\ref{maxH}) directly,
\begin{equation*}
\frac{\partial \call_a(S;H^{+}_a)}{\partial H^{+}_a}=-\gamma^{+}_a  \left(\frac{S}{H^{\pm}_a}\right)^{\gamma^{\pm}_{a}}\left[\frac{\gamma^{\pm}_a-1}{\gamma^{\pm}_a} \frac{\delta_a}{\lambda +\delta_a} -\frac{K}{H^{+}_a}\right],
\end{equation*}
what leads to the following algebraic expression for $H^+_a$:
\begin{equation*}
H^+_a = \frac{\gamma^{+}_a(1+\lambda/\delta_a)}{\gamma^{+}_a-1} K.
\end{equation*}
Note that this expression for $H^+_a$ leads to $\call_a<S$. This result has its origins in the uneven rights that option holders and stock owners have before the change due to the dividend payment. Remember that after the change $\call_b=S$, so the call holder must face the following dilemma: if the option is not exercised there is still a chance for the sudden revaluation of the option price, since $\call_b>\call_a$, but at the same time, the longer he/she waits the higher is the amount of dividends he/she will never receive. The optimal exercise price $H^+_a$ comes from a trade-off between these two competing reasons. Thus the difference disappears if we consider the limit $\delta_a\rightarrow 0$: we can see how we recover $H^{+}_a\rightarrow \infty$, the option is never exercised also in this case, and $\call_a\rightarrow S$ as well. Therefore if the share pays no dividend at all, American call options quote as its underlying even if the volatility changes.

\begin{figure}[hbt] 
{\hfill
\includegraphics[width=0.80\textwidth,keepaspectratio=true]{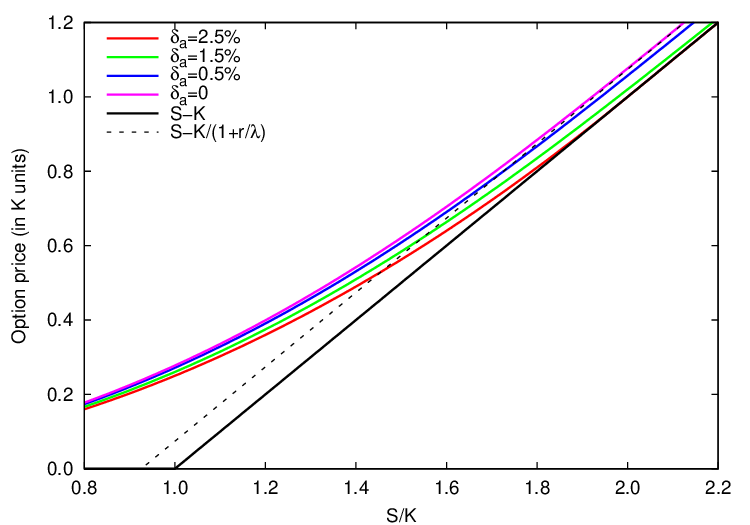} 
}
\caption{Option prices for perpetual vanilla calls under dividend risk.
We represent the price of the option, in terms of the moneyness ($S/K$), for different values of $\delta_a$. We observe how both the optimal exercise price and the option price itself increase as $\delta_a$ tends to zero. When $\delta_a=0$ the option price never touches the pay-off function,  depicted in a solid black line. We have used the following values for the parameters: $r=4\%$, $\sigma_a=\sigma_b=10\%$, $\delta_b=2.5\%$ and $\lambda^{-1}=2$ years.}
\label{figCa2d0}
\end{figure}

Let us assume next that $\delta_a=0$ but with $\delta_b>0$, the stock can start suddenly to pay dividends. Since $\beta^+_b>\beta^+_a$ the appropriate pricing expressions follows from~(\ref{Vcaputa2}) and~(\ref{Vcaputa3}), and $H^+_a/H^+_b$ should be a zero of the auxiliary function $\HH^+(\chi)$. The first point that deserves attention is the fact that in this case we have $C^+=0$, $D^+>0$ and, as a consequence, $\HH^+(\chi)$ is strictly positive. 
Then the only way in which one can satisfy the maximal condition~(\ref{maxH}) is by demanding that $H^+_a\rightarrow \infty$. This conclusion must hold regardless the value of the rest of the parameters in general, and  $\beta^+_b>1$ in particular. The validity of this result was already conjectured in~\cite{MM07}, but no prove was given there. Note that in this case the price of the call differs from the underlying, $\call_a<S$, due to the existence of the post-change scenario, $\call_b<S$:
\begin{eqnarray*}
\call_a(S)=
\frac{\lambda}{\lambda+r}\frac{\beta^+_b}{\gamma^+_a-\gamma^-_a}\frac{\gamma^-_a}{(\gamma^+_a-1)(\gamma^+_a-\beta^+_b)} K\left(\frac{S}{H^+_b}\right)^{\gamma^+_a}\\
+(H^{+}_b - K)\frac{\lambda}{\lambda+\ell^{+}} \left(\frac{S}{H^{+}_b}\right)^{\beta^{+}_b}, \quad (S\leqslant H^{+}_b),
\end{eqnarray*}
and
\begin{eqnarray*}
\call_a(S)=\frac{\lambda}{\lambda+r}\frac{\beta^+_b}{\gamma^+_a-\gamma^-_a}\frac{\gamma^+_a}{(1-\gamma^-_a)(\beta^+_b-\gamma^-_a)} K\left(\frac{S}{H^+_b}\right)^{\gamma^-_a}\\
+S - \frac{\lambda}{\lambda+r}K, \quad ( H^{+}_b<S<\infty).
\end{eqnarray*}
Only in the limit $\lambda \rightarrow 0$, the solution $\call_a=S$ is recovered. 
Figure~\ref{figCa2d0} shows how $H^+_a$ increases when one considers lower and lower values for $\delta_a$, until the limit $\delta_a=0$ is attained. In this case the price never intersects the pay-off function.

\begin{figure}[hbt] 
{\hfill
\includegraphics[width=0.80\textwidth,keepaspectratio=true]{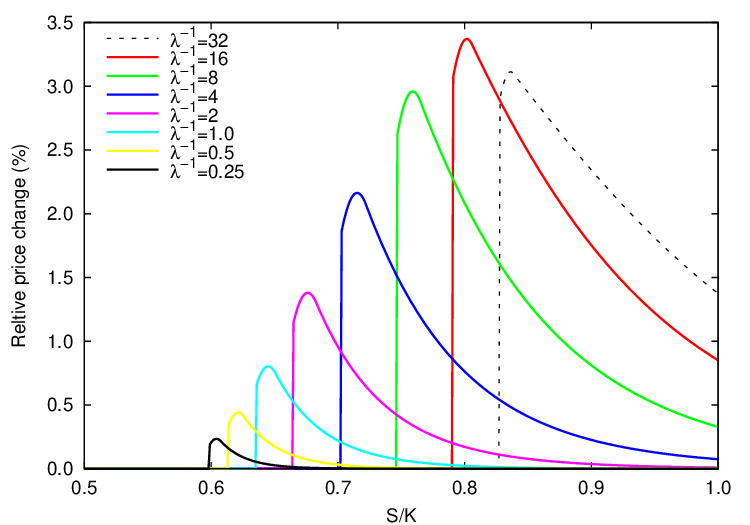} 
}
\caption{Relative price change for a perpetual vanilla put under volatility risk.
We represent the price of the option, in terms of the moneyness ($S/K$), for different values of $\lambda^{-1}$, expressed in years. We analyse here the difference between expressions~(\ref{Vcaputa1}) and~(\ref{Vcaputa2}) when a sudden increment in the volatility of the stock is presumed. We have used the following values for the parameters: $r=4\%$, $\delta_a=\delta_b=0$, $\sigma_a=10\%$ and $\sigma_b=25\%$. Note how the right solution dictates a higher price.}
\label{figPrel}
\end{figure}

Another aspect of the previous solutions that it is worthwhile devoting attention to is the behaviour of those expressions out of their range of validity. As we have pointed out in the previous Section, and partially shown in the Appendix, equation $\HH^\pm(\chi^{\pm}_0)=0$ has a unique solution in the interval $1<\chi^{\pm1}<\infty$ when $|\beta^{\pm}_a|< |\beta^{\pm}_b|$. The fact is that if one considers arbitrary relative values for $\beta^{\pm}_{\ab}$ and expands the domain of $\HH^\pm(\chi)$ to the whole positive real axis, it {\it always\/} presents a single zero, because $\lim_{\chi\rightarrow 0} \HH^\pm(\chi)<0$, $\lim_{\chi\rightarrow \infty} \HH^\pm(\chi)>0$, and the function is either monotonous or has a single local extreme. Therefore one can compute $\hat{H}^{\pm}_a=\hat{\chi}^{\pm}_0 H^{\pm}_b$, $\HH^\pm(\hat{\chi}^{\pm}_0)=0$, when $|\beta^{\pm}_a|\geqslant|\beta^{\pm}_b|$, and compare it with the right value $H^{\pm}_a=\eta^{\pm}_0 H^{\pm}_b$. The same kind of analysis can be performed on the two associated pricing expressions: $\caput_a$, the correct one in the present circumstances, found in~(\ref{Vcaputa1}), and the heuristic one $\hat\caput_a$, obtained by using formula~(\ref{Vcaputa2}). Note that solution~(\ref{Vcaputa3}) does not apply in any case. This is not a mere academic exercise, because both expressions solve the {\it same\/} equation, namely~(\ref{odeCalla1}) for call options, and $\hat{H}^{\pm}_a$ still obeys to a maximization principle. The divergence emerges as a consequence of the different boundary conditions to be satisfied: in particular it is no longer true that $\hat\caput_a(S=\hat{H}^{\pm}_a)=\pm(\hat{H}^{\pm}_a-K)$. Since this condition is deduced from arguments on optimality, one can consider the heuristic expression as a non-optimal solution.  In Figure~\ref{figPrel} we can observe the relative difference between pricing expressions, $(\putt_a-\hat\putt_a)/\putt_a$, for a perpetual put option in a market model with $\beta^{-}_a=-8.0$ and $\beta^{-}_b=-1.28$.  
The heuristic solution underprices the put in a quantity that reduces for large and small values of $S$ and $\lambda$, because then the mismatch in the boundary reduces its relevance. In any case, the error introduced by the heuristic pricing expression in the terms analysed is relatively small and bounded. In fact, as we show below with a different example, if we simply superimposed the two prices in a single plot it would seem that they just collapse. Besides $\putt_a\geqslant\hat\putt_a$ we have also $H^{-}_a\geqslant\hat{H}^{-}_a$. This is tenable because the sudden change in the heuristic solution at $S=\hat{H}^{-}_a$, clearly noticeable in Figure~\ref{figPrel}. $\hat\putt_a$ also crosses the pay-off function at $S>\hat{H}^{-}_a$, what reinforces the heuristic character of the solution. 

\begin{figure}[hbt] 
{\hfill
\includegraphics[width=0.80\textwidth,keepaspectratio=true]{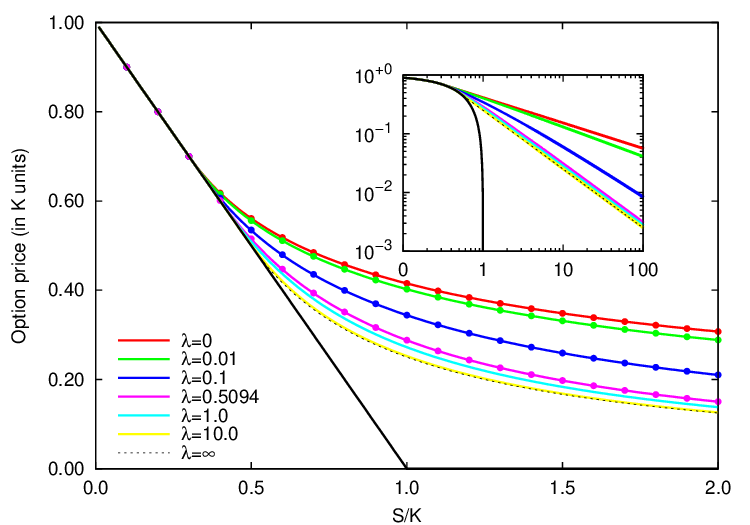} 
}
\caption{Option prices for a perpetual vanilla put under volatility risk.
We represent the price of the option, in terms of the moneyness ($S/K$), for different values of $\lambda$, expressed in years$^{-1}$, when one awaits a severe reduction in the volatility of the stock. We observe how right prices (solid lines) coincide optically with heuristic ones (dots) up to the threshold $\bar{\lambda}\approx 0.5094$ is attained, even for large stock prices |see the inset. For higher values of $\lambda$ we have only the correct solution. We have used the following values for the rest of parameters: $r=4\%$, $\delta_a=\delta_b=1.75\%$, $\sigma_a=40\%$ and $\sigma_b=25\%$.}
\label{fig3OK}
\end{figure}

Let us consider now the opposite situation in which one tries to use~(\ref{Vcaputa1}) instead of expressions~(\ref{Vcaputa2})-(\ref{Vcaputa3}) when $|\beta^{\pm}_a| < |\beta^{\pm}_b|$. The immediate conclusion is that in this case~(\ref{Vcaputa1}) does not solve the appropriate differential equation, because $\caput_b$ |which plays the role of an external interaction| is not properly settled, but right boundary conditions are to be satisfied eventually. This may result into incompatibility. Note that $\lim_{\eta\rightarrow 0} \GG^\pm(\eta)\lessgtr 0$, and  $\lim_{\eta\rightarrow \infty} \GG^\pm(\eta)\lessgtr 0$ as well. This implies here that $\GG^\pm(\eta)$ has either two zeros or none. If $|\beta^{\pm}_a| \geqslant |\beta^{\pm}_b|$ we have stated that the problem has always a univocal solution, but when  $|\beta^{\pm}_a| < |\beta^{\pm}_b|$ it is possible that no extremal solution can be defined. This problem was already detected and depicted in~\cite{MM07}. In Figure~\ref{fig3OK} we observe how, for a given market conditions, no heuristic solution can be found beyond a critical value of $\lambda$,  $\bar{\lambda}\approx 0.5094$. Once again the value of the heuristic optimal exercise price is smaller than ${H}^{-}_a$, and since the heuristic solution fulfils all the boundary conditions, it overprices the value of the option.  

\begin{figure}[hbt] 
{\hfill
\includegraphics[width=0.80\textwidth,keepaspectratio=true]{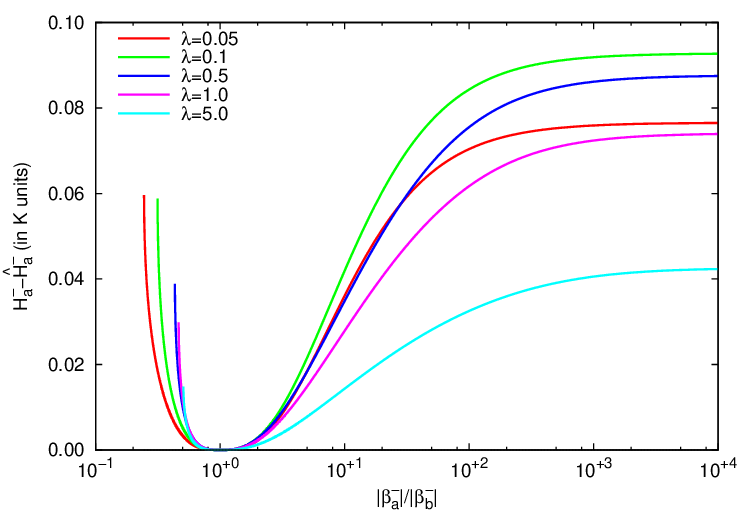} 
}
\caption{Optimal exercise price difference. We represent ${H}^{-}_a-\hat{H}^{-}_a$ as a function of the ``order parameter" $|\beta^{-}_a|/|\beta^{-}_b|$, for different values of $\lambda$, expressed in years$^{-1}$. We observe how the true value is always maximal, the difference is bounded, and $\hat{H}^{-}_a$ ends in a point where its derivative is infinite. In the confection of the plot we have varied $\sigma_a$ and left unchanged the rest of parameters: $r=4\%$, $\delta_a=\delta_b=1.75\%$, and $\sigma_b=25\%$.}
\label{figHH}
\end{figure}

Apart from the surprising accuracy of the heuristic solutions, the main conclusion of the previous discussion is that the right $H^{\pm}_a$ is extremal among different configurations, Figure~\ref{figHH}. Therefore one can optionally interpret that a solution cease to be valid {\it because\/} it conveys to an exercise price which is not as optimal as it could be, like stable and metastable equilibrium states in thermodynamics.

Note finally that when $\beta^{\pm}_a=\beta^{\pm}_b$, not only $H^{\pm}_a=H^{\pm}_b$ but also $\caput_a=\caput_b$, no matter the values that the market parameters take individually.  
\section{Conclusions}
\label{Last}
In this article we have obtained the complete set of analytic pricing expressions for perpetual American vanilla options within a market model that may present a single change of regime in both the dividend-payment rate and the volatility level of the underlying stock. We further assume that the value of these parameters after the change are known quantities. The problem is relevant from the point of view of mathematical finance and shows interesting connections with physics.

The most exciting financial result is perhaps that the price of a perpetual American vanilla call when the underlying does not pay dividends may differ from the spot price of the stock, regardless the fact that the option will be never early exercised as long as the market conditions remain fixed. The risk associated to the fact that this non-dividend policy can be eventually revised in the future reduces the value of the option.     

On the realms of physics one finds certain parallelisms between some properties shown by this model and phase transitions in thermodynamics: when one analyses the issue before the regime change, two different formulas are found for both the price of the option and the optimal exercise boundary, and the appropriate expression to be used is elucidated on the basis of the value of a some sort of ``order parameter". 

This analogy with thermodynamics extends to the mutual relationship between the two available exercise prices. Like state functions in thermodynamics, exercise prices characterize equilibrium states in the sense that they are solutions to extreme value problems. When two possible values for a function of state coexist, the one that is smaller (or larger, depending on the state quantity analysed) is seen as the stable solution, whereas the alternative solution is named as metastable. In our case this kind of behaviour is observed, and one can trace the reason back to the same origin: each solution  reflects different environmental conditions, and only one of them incorporates the actual constraints. And as in physical problems, the metastable solution may become eventually unstable and just disappear. We have shown how this phenomenon was on the basis of the spurious bizarre results reported in~\cite{MM07}.

Finally, we have illustrated the major results enumerated above with several graphical examples corresponding to practical market situations. 

\ack The author acknowledges partial support from the former Spanish {\it Ministerio de Educaci\'on y Ciencia\/} under contract no. FIS2006-05204-E, from the {\it Generalitat de Catalunya\/} under contract no. 2005 SGR-00515, and from the {\it Junta de Castilla y Le\'on\/} under contract no. SA034A08.
\appendix
\section{Detailed computations}
\label{AA}
Let us consider the case of call options in detail. The equation to be solved if the regime has changed is~(\ref{Xi_b}):
\begin{equation*}
\frac{1}{2} \sigma^2_b S^2 \frac{\dd^2 \call_b}{\dd S^2} +(r-\delta_b) S\frac{\dd \call_b}{\dd S} -r \call_b=0, \quad (S\leqslant H^{+}_b), 
%\label{odePb}
\end{equation*}
whose general solution reads
\begin{equation*}
\call_b(S;H^{+}_b)=\kappa_{1} S^{\beta^{+}_b}+\kappa_0 S^{\beta^{-}_b},
\end {equation*}
where $\beta^{\pm}_b$ first appeared in~(\ref{betapm}). The applicable boundary conditions are three in this case:
\begin{eqnarray}
\lim_{S\rightarrow 0}\call_b &=&0, \label{vb_1}\\
\lim_{S\rightarrow H^{+}_b}\call_b&=&H^{+}_b-K, \label{vb_2}\\
\lim_{S\rightarrow H^{+}_b}\frac{\dd \call_b}{\dd S}&=&1. \label{vb_3}
\end{eqnarray}
Since $\beta^{\pm}_{b}\gtrless 0$ we have that $\kappa_0=0$ by virtue of~(\ref{vb_1}). $\kappa_1$ is obtained after (\ref{vb_2}),
\begin{equation*}
\kappa_1=(H^{+}_b - K)\left(\frac{1}{H^{+}_b}\right)^{\beta^{+}_b},
\end{equation*}
and therefore~\cite{BAW87}: 
\begin{equation}
\call_b=(H^{+}_b - K) \left(\frac{S}{H^{+}_b}\right)^{\beta^{+}_b},\quad (S\leqslant H^{+}_b) .
\label{call_b}
\end{equation}
The value of $H^{+}_b$ emerges from~(\ref{vb_3}):
\begin{equation}
H^{+}_b = \frac{\beta^{+}_b}{\beta^{+}_b - 1} K. \label{Hplusb}
\end{equation}
It can be proven that $\beta^{+}_b \geqslant 1$, and that $\beta^{+}_b = 1$ is equivalent to $\delta_b=0$. Therefore we will have $K<H^{+}_b<\infty$ for $\beta^{+}_b > 1$, and for $\beta^{+}_b = 1$  the value of $H^{+}_b$ diverges: there is no optimal boundary, the option is never exercised and quotes as the underlying stock, 
\begin{equation*}
\lim_{\beta^+_b\rightarrow 1}\call_b=S.
\end{equation*}
Note that~(\ref{call_b}) is valid for $S\leqslant H^{+}_b$, whereas the value of the call for $S> H^{+}_b$ is just $S-K$. Some remarks about this statement can become very revealing for a better understanding of future developments. The exercise price of a call marks the threshold above which it is better to execute the option rather than to keep it alive. However, this does not ensure that one will be able to exercise the option exactly when $S=H^{+}_b$. Let us think, for instance, in the case of a market with quantified quotes,~\footnote{In fact, most of real markets behave in that way, because they demand a minimum variation to change a price. This tick size may be as small as a cent, but this does not invalidate the argument.} and observe how in general the {\em executed\/} price will exceed the optimal exercise price unless it coincides with one of the admissible prices. In our case, the process $S_t$ is continuous but the  optimal exercise price is not, it changes from $H^{+}_a$ to $H^{+}_b$, what can end in the same result if $H^{+}_a>H^{+}_b$. 

Let us consider now the call price before the change, the one we are primarily interested in. The differential equation is
\begin{equation*}
\frac{1}{2} \sigma^2_a S^2 \frac{\dd^2 \call_a}{\dd S^2} +(r-\delta_a) S\frac{\dd \call_a}{\dd S} -(r+\lambda) \call_a=-\lambda \call_b, \quad (S\leqslant H^{+}_a), 
%\label{odePb}
\end{equation*}
and therefore one must choose the right expression for $\call_b$. Two well different scenarios emerges, depending on the relative values of $H^{+}_a$ and $H^{+}_b$. If $H^{+}_a\leqslant H^{+}_b$ we will have a single equation to solve
\begin{equation}
\frac{1}{2} \sigma^2_a S^2 \frac{\dd^2 \call_a}{\dd S^2} +(r-\delta_a) S\frac{\dd \call_a}{\dd S} -(r+\lambda) \call_a=-\lambda (H^{+}_b - K) \left(\frac{S}{H^{+}_b}\right)^{\beta^{+}_b} , 
\label{odeCalla1}
\end{equation}
for all $S\leqslant H^{+}_a$, whereas when $H^{+}_a>H^{+}_b$, a set of two different equations appears, equation (\ref{odeCalla1}) itself for $S\leqslant H^{+}_b$, and
\begin{equation}
\frac{1}{2} \sigma^2_a S^2 \frac{\dd^2 \call_a}{\dd S^2} +(r-\delta_a) S\frac{\dd \call_a}{\dd S} -(r+\lambda) \call_a=-\lambda (S - K),
\label{odeCalla22}
\end{equation}
for $H^{+}_b<S\leqslant H^{+}_a$. The point is that the value of $H^{+}_a$ is obtained after the resolution of the problem. Therefore we will start our analysis by assuming that  $H^{+}_a\leqslant H^{+}_b$. The general solution of~(\ref{odeCalla1}) is
\begin{equation}
\call_a=\kappa_3 S^{\gamma^{+}_a}+\kappa_2 S^{\gamma^{-}_a} +\frac{\lambda}{\lambda+\ell^{+}}(H^{+}_b-K)\left(\frac{S}{H^{+}_b}\right)^{\beta^{+}_b}, \ (S\leqslant H^{+}_a\leqslant H^{+}_b),
\label{odeCalla1Sol}
\end {equation}
where $\gamma^{\pm}_a$ are defined in~(\ref{gamma}), $\ell^+$ in~(\ref{ell}), and we are assuming for the moment that $\gamma^{+}_a\neq\beta^{+}_b$.
Now we can impose the boundary conditions
\begin{eqnarray}
\lim_{S\rightarrow 0}\call_a&=&0,\label{va1_1}\\
\lim_{S\rightarrow H^{+}_a}\call_a&=&H^{+}_a-K,\label{va1_2}\\
\lim_{S\rightarrow H^{+}_a}\frac{\dd \call_a}{\dd S}&=&1.\label{va1_3}
\end{eqnarray}
From~(\ref{va1_1}) we deduce $\kappa_2=0$, and~(\ref{va1_2}) leads to
\begin{equation*}
\kappa_3=\left[H^{+}_a-K-\frac{\lambda}{\lambda+\ell^{+}}(H^{+}_b-K)\left(\frac{H^{+}_a}{H^{+}_b}\right)^{\beta^{+}_b}\right]\left(\frac{1}{H^{+}_a}\right)^{\gamma^{+}_a},
\end{equation*}
and then~\cite{MM07}:
\begin{eqnarray}
\call_a =
(H^{+}_a - K)\left(\frac{S}{H^{+}_a}\right)^{\gamma^{+}_{a}} \nonumber \\  +(H^{+}_b - K)\frac{\lambda}{\lambda+\ell^{+}} \left[\left(\frac{S}{H^{+}_b}\right)^{\beta^{+}_b} -\left(\frac{H^{+}_a}{H^{+}_b}\right)^{\beta^{+}_b} \left(\frac{S}{H^{+}_a}\right)^{\gamma^{+}_{a}} \right], (S\leqslant H^{+}_a).
\label{Vplusa1}
\end{eqnarray}
The value of $H^{+}_a$ must fulfil the transcendental equation that follows from the smoothness condition~(\ref{va1_3}),
\begin{equation*}
\gamma^{+}_a-1-\frac{2 \lambda}{\sigma^2_a \beta^+_b(\beta^+_b-\gamma^-_a)} \left(\frac{H^{+}_a}{H^{+}_b}\right)^{\beta^{+}_b-1}-\gamma^{+}_a \frac{K}{H^{+}_a}=0, %\label{Hplusa1}
\end{equation*}
where $H^+_b$ was introduced in~(\ref{Hplusb}). 
When $\beta^+_b\neq 1$ we can define the auxiliary function $\GG^{+}(\eta)$, 
\begin{equation*}
\GG^{+}(\eta) = A^{+} \eta^{1-\beta^{+}_b} - B^{+} -\eta^{-\beta^{+}_b}, 
%\label{Gpm}
\end{equation*}
with $A^{+}$ and $B^{+}$ bounded and positive constants |see~(\ref{cA}) and~(\ref{cB}) in the main text for their definitions.
The analysis of this function is relevant in our framework because $H^+_a$ fulfils $\GG^{+}(H^+_a/H^+_b)=0$. 
Since we have assumed that $H^{+}_a\leqslant H^{+}_b$, we are interested in studying the properties of $\GG^{+}(\eta)$ in the range $0<\eta\leqslant 1$. The values of the function in the interval endings are $\lim_{\eta \rightarrow 0}\GG^{+}(\eta)=-\infty$ and  $\GG^{+}(1)= \ell^+ B^{+}/\lambda$. 
$\GG^{+}(\eta)$ shows a single extremal value located outside the segment $\eta\in (0,1]$:
\begin{equation*}
\eta_M=\frac{\gamma^{+}_a}{\gamma^{+}_a-1}>1, 
\end{equation*}
which is a maximum since
\begin{equation*}
\lim_{\eta \rightarrow \eta_M}\frac{\dd^2 \GG^{+}}{\dd \eta^2}=-\beta^+_b \left(\eta_M\right)^{-\beta^+_b-2}<0. 
\end{equation*}
The above properties compel equation $\GG^{+}(\eta_0^+)=0$ to have at the most one solution for which $0<\eta_0^+\leqslant 1$. The necessary and sufficient condition for the existence of such solution is 
\begin{equation*}
\GG^{+}(1)\geqslant 0 \Leftrightarrow \ell^+\geqslant 0 \Leftrightarrow\beta^+_a\geqslant\beta^+_b.
\end{equation*}
Therefore, condition $H^{+}_a\leqslant H^{+}_b$ grants that $\beta^+_a\geqslant\beta^+_b$ and, incidentally, that  
$\gamma^+_a>\beta^+_b$ is satisfied in~(\ref{Vplusa1}). 

In fact, we can show how $H^+_a>H^+_b$ leads to $\beta^+_a<\beta^+_b$ as well.
Let us assume that $H^+_a>H^+_b$: in this case we have to solve~(\ref{odeCalla1}) and~(\ref{odeCalla22}). The general solutions if $\gamma^+_a\neq\beta^+_b$ are:
\begin{equation*}
\call_a=\kappa_5 S^{\gamma^{+}_a}+\kappa_4 S^{\gamma^{-}_a} +\frac{\lambda}{\lambda+\ell^{+}}(H^{+}_b-K)\left(\frac{S}{H^{+}_b}\right)^{\beta^{+}_b}, \quad (S\leqslant H^{+}_b),
%\label{odeCalla21Sol}
\end{equation*}
and 
\begin{equation*}
\call_a=\kappa_7 S^{\gamma^{+}_a}+\kappa_6 S^{\gamma^{-}_a} +\frac{\lambda}{\lambda+\delta_a}S-\frac{\lambda}{\lambda+r} K, \quad (H^{+}_b<S\leqslant H^{+}_a),
%\label{odeCalla22Sol}
\end{equation*}
with the five boundary conditions to be satisfied listed below:
\begin{eqnarray}
\lim_{S\rightarrow 0}\call_a&=&0,\label{va2_1}\\
\lim_{S\uparrow H^{+}_b}\call_a&=&\lim_{S\downarrow H^{+}_b}\call_a,\label{va2_2}\\
\lim_{S\uparrow H^{+}_b}\frac{\dd \call_a}{\dd S}&=&\lim_{S\downarrow H^{+}_b}\frac{\dd \call_a}{\dd S},\label{va2_3}\\
\lim_{S\rightarrow H^{+}_a}\call_a&=&H^{+}_a-K,\label{va2_4}\\
\lim_{S\rightarrow H^{+}_a}\frac{\dd \call_a}{\dd S}&=&1.\label{va2_5}
\end{eqnarray}
After imposing constraints~(\ref{va2_1})-(\ref{va2_4}) one gets
\begin{eqnarray*}
\call_a=\Bigg[\frac{\delta_a}{\lambda+\delta_a}H^+_a - \frac{r}{\lambda+r}K\Bigg]\left(\frac{S}{H^+_a}\right)^{\gamma^+_a}\\
+\frac{\lambda}{\lambda+r}\frac{\beta^+_b}{\gamma^+_a-\gamma^-_a} K\Bigg[\frac{\gamma^-_a}{(\gamma^+_a-1)(\gamma^+_a-\beta^+_b)}\\-\frac{\gamma^+_a}{(1-\gamma^-_a)(\beta^+_b-\gamma^-_a)}\left(\frac{H^+_b}{H^+_a}\right)^{\gamma^+_a-\gamma^-_a}\Bigg]\left(\frac{S}{H^+_b}\right)^{\gamma^+_a}\\
+(H^{+}_b - K)\frac{\lambda}{\lambda+\ell^{+}} \left(\frac{S}{H^{+}_b}\right)^{\beta^{+}_b}, \quad (S\leqslant H^{+}_b),
\end{eqnarray*}
and
\begin{eqnarray*}
\call_a=\left[\frac{\delta_a}{\lambda+\delta_a}H^+_a - \frac{r}{\lambda+r}K\right]\left(\frac{S}{H^+_a}\right)^{\gamma^+_a}\\
+\frac{\lambda}{\lambda+r}\frac{\beta^+_b\gamma^+_a}{(\gamma^+_a-\gamma^-_a)(1-\gamma^-_a)(\beta^+_b-\gamma^-_a)} K\Bigg[\left(\frac{S}{H^+_b}\right)^{\gamma^-_a}-\left(\frac{H^+_a}{H^+_b}\right)^{\gamma^-_a}\left(\frac{S}{H^+_a}\right)^{\gamma^+_a}\Bigg]\\
+\frac{\lambda}{\lambda+\delta_a}S - \frac{\lambda}{\lambda+r}K, \quad ( H^{+}_b<S\leqslant H^{+}_a).
\end{eqnarray*}
Condition~(\ref{va2_5}) leads to a new transcendental equation:
\begin{eqnarray*}
\frac{\lambda(\beta^+_b-1)}{(1-\gamma^-_a)(\beta^+_b-\gamma^-_a)} \left(\frac{H^+_a}{H^+_b}\right)^{\gamma^-_a-1} + r  \frac{K}{ H^+_a}+\frac{\gamma^-_a}{1-\gamma^-_a}\delta_a=0,
\end{eqnarray*}
and one can define a second auxiliary function in order to analyse the problem:
\begin{eqnarray*}
\HH^+(\chi)= \chi^{-\gamma^-_a}-C^+ \chi^{1-\gamma^-_a}+D^+,
\end{eqnarray*}
since $\HH^+(H^+_a/H^+_b)=0$. As before $C^+$ and $D^+$ are bounded and positive constants,  defined in the main text, equations~(\ref{cC}) and~(\ref{cD}).
The values at the extremes of the interval of interest are $\lim_{\chi \rightarrow \infty}\HH^{+}(\chi)=-\infty$ and 
\begin{equation*}
\lim_{\chi \rightarrow 1} \HH^{+}(\chi)=\frac{\gamma^-_a}{(\beta^+_b-\gamma^-_a)(\beta^{+}_b-1)} \frac{\ell^+}{r}>0,
\end{equation*}
provided that $\beta^+_a<\beta^+_b$. When $\delta_a\neq0$ function  ${\cal H}^+(\chi)$ has a maximum at $\chi_M$,
\begin{equation*}
\chi_M=\frac{\beta^{+}_b}{\beta^{+}_b-1}\frac{\delta_a}{r},
\end{equation*}
since
\begin{equation*}
\lim_{\chi \rightarrow \chi_M}\frac{\dd^2 \HH^{+}}{\dd \chi^2}=\gamma^-_a \left(\chi_M\right)^{-\gamma^-_a-2}<0, 
\end{equation*}
whereas ${\cal H}^+(\chi)$ is a decreasing function if $\delta_a=0$. In any case, one can conclude that
$\HH^{+}(\chi)$ has a single zero $\chi^+_0\in (1,\infty)$, irrespective of the maximum location which, if exists, may be placed either inside or outside this region.

The resolution scheme for perpetual American vanilla puts is very similar. We have decided not to include it here for brevity reasons.

\end{document}